\documentclass[%
reprint,
amsmath,amssymb,
aps,
pra,
superscriptaddress,
]{revtex4-1}

\usepackage{ulem}
\usepackage{mathtools}
\usepackage{braket}
\usepackage{graphicx}
\usepackage{dcolumn}
\usepackage{bm}
\usepackage{hyperref}
\usepackage{color}
\hypersetup{colorlinks=true, citecolor = blue, linkcolor = blue}

\begin{document}

\preprint{APS/123-QED}

\title{Entanglement distribution between quantum repeater nodes
\\with an absorptive type memory}

\author{Daisuke Yoshida}
\email{yoshida-daisuke-mv@ynu.jp}
\affiliation{
Yokohama National University, 79-5 Tokiwadai, Hodogaya, Yokohama 240-8501, Japan
}
\author{Kazuya Niizeki}%
\affiliation{
Yokohama National University, 79-5 Tokiwadai, Hodogaya, Yokohama 240-8501, Japan
}
\author{Shuhei Tamura}
\affiliation{
Yokohama National University, 79-5 Tokiwadai, Hodogaya, Yokohama 240-8501, Japan
}
\author{Tomoyuki Horikiri}
\affiliation{
Yokohama National University, 79-5 Tokiwadai, Hodogaya, Yokohama 240-8501, Japan
}
\affiliation{
JST, PRESTO, 4-1-8 Honcho, Kawaguchi, Saitama, 332-0012, Japan
}

\date{\today}

\begin{abstract}
Quantum repeaters, which are indispensable for long-distance quantum communication, are necessary for extending the entanglement 
from short distance to long distance; however, high-rate entanglement distribution, even between adjacent repeater nodes, has not been realized.
In a recent work by C. Jones, et al., New J. Phys. 18, 083015 (2016), the entanglement distribution rate between adjacent repeater nodes was calculated for a plurality of quantum dots, nitrogen-vacancy centers in diamond, and trapped ions adopted as quantum memories inside the repeater nodes. 
Considering practical use, arranging a plurality of quantum memories becomes so difficult with the state-of-the art technology.
It is desirable that high-rate entanglement distribution is realized with as few memory crystals as possible.

Here we propose new entanglement distribution scheme \color{black}
with one quantum memory based on the atomic frequency comb which enables temporal multimode operation with one crystal. The adopted absorptive type quantum memory degrades the difficulty of multimode operation compared with previously investigated quantum memories directly generating spin-photon entanglement.
It is shown that the present scheme improves the distribution rate by nearly two orders of magnitude compared with the result in C. Jones, et al., New J. Phys. 18, 083015 (2016) and the experimental implementation is close by utilizing state-of-the-art technology.
\color{black}

\end{abstract}

\pacs{Valid PACS appear here}
\maketitle


\newcommand*{\equlab}[1]{\label{equ:#1}}
\newcommand*{\equref}[1]{Eq.~\ref{equ:#1}}
\newcommand*{\seclab}[1]{\label{sec:#1}}
\newcommand*{\secref}[1]{Section~\ref{sec:#1}}
\newcommand*{\figlab}[1]{\label{fig:#1}}
\newcommand*{\figref}[1]{Fig.~\ref{fig:#1}}
\newcommand*{\tablab}[1]{\label{tab:#1}}
\newcommand*{\tabref}[1]{Table~\ref{tab:#1}}

\section{Introduction}\seclab{intro}

The loss of photons in an optical fiber exponentially increases with distance. 
Therefore, quantum repeaters~\cite{Briegel1998, Duan2001, Jones2016} are necessary to realize a secure, long-distance entanglement distribution, which is vital for various quantum communication technologies including quantum key distribution~\cite{Bennett1988, Ekert1991}, quantum teleportation~\cite{Bennett1993}, and blind quantum computation~\cite{AnneBroadbentElhamKashefi2009}. 
In recent years, various quantum repeater protocols using quantum memories have been proposed~\cite{Sangouard2011} and research was conducted on several materials, including semiconductor quantum dots (QDs)~\cite{Greve2012}, nitrogen-vacancy (NV) centers in diamond~\cite{Togan2010, Kosaka2015, Yang2016, Bernien2013}, trapped ions~\cite{Blinov2004}, and atomic ensembles in a gas or solid~\cite{Sangouard2011}. 
Efforts have been made to increase the entanglement distribution rate with the repeater protocols for practical use. In order to improve the rate
\color{black}
and the feasibility 
\color{black}, the optimization of the arrangement of each element in the repeater system, including the quantum memory and entangled photon source (EPS), is necessary.
 
In this study, we analyze the entanglement distribution rate between adjacent repeater nodes using two types of quantum communication schemes \color{black}in consideration of the feasibility\color{black}. 
In Ref.~\cite{Jones2016}, the entanglement distribution rates by several kinds of quantum memories and arrangements of each element were compared. Specifically, the entanglement distribution rates were compared using three schemes: meet-in-the-middle (MM)~\cite{Simon2003, Duan2001, Feng2003}, sender-receiver (SR)~\cite{Munro2010}, and midpoint source (MS)~\cite{Jones2013}. In addition, three kinds of quantum memories including QDs, NV centers in diamond, and trapped ions were compared when the MM and MS were performed.

In Ref.~\cite{Jones2016}, quantum memories based on an atomic ensemble were not adopted.
In recent years, a technique called the atomic frequency comb (AFC) using an atomic ensemble (especially rare-earth ion ensembles in a solid) was proposed and developed as a quantum memory~\cite{Afzelius2009}. 
The main feature of this memory is temporal multiplexing \color{black}with one crystal, \color{black}and the absorption and reemission efficiency can be 100\% in principle.

\color{black}
Here, we will briefly introduce the idea of AFC quantum memory~\cite{Afzelius2009}. A rare-earth-ion-doped solid (REIDS) is assumed to be the memory material. In AFC quantum memory, three energy levels, $ \ket{e}, \ket{g}, \ket{s} $, are utilized. $\ket{e}$ is the excited state, $\ket{g}$ is the ground state, and $\ket{s}$ is one of the hyperfine or Zeeman sublevels. 
A certain spectral range inside a wide inhomogeneous broadening is made to be transparent using a hole-burning technique with strong pumping by an intense laser. In the range called the spectral pit, absorption lines corresponding to the transition of $\ket{g}$ to $\ket{e}$ are generated with an interval of $\Delta$ by a control laser. When a single photon is absorbed by the memory material, it excites one ion in a large ensemble. 
The state can be described by the following equation:
\begin{equation}
\ket{\psi} = \sum^{N}_{j=1}c_je^{i\delta_jt}e^{-ikz_j}\ket{g_1\cdots e_j ,\cdots g_{N}}, \equlab{addeq1}
\end{equation}
where $z_j$ is the position of the ion $j$, $k$ is the wave number of the light field, $\delta_j$ is the detuning of the ion with respect to the laser frequency, and the amplitude $c_j$ depends on the frequency and spatial position of the particular ion $j$.
After absorbing a single photon, the phase factor of the collective state experiences dephasing. However, after the time of $2\pi / \Delta$, all the terms are rephased and single-photon emission as well as a transition to $\ket{g}$ occur.
To obtain on-demand memory time, the excitation at $\ket{e}$ can be transferred to $\ket{s}$ by Rabi oscillation using a $\pi$ pulse. In $\ket{s}$, long-term preservation is possible using the spin-echo method~\cite{Zhong2015}. By using a $\pi$ pulse again, the transition from $\ket{s}$ to $\ket{e}$ is achieved and on-demand memory time becomes possible. In the AFC scheme, the absorption and retrieval efficiency can, in principle, reach 100\% by optimizing the doping rate of rare-earth additives and the finesse of the comb~\cite{Afzelius2009}. In addition, contrary to previous proposals including electromagnetically induced transparency (EIT)~\cite{Fleischhauer2000} and controlled reversible inhomogeneous broadening (CRIB)~\cite{Moiseev2001}, which also use an atomic ensemble, it is possible to increase the number of modes without reducing the efficiency while maintaining the optical depth. The number of multiplexing temporal modes is proportional to the number of combs and is possible in hundreds of modes.~\cite{Afzelius2009} 
\color{black}

\color{black}
Therefore, it is worth studying a possibility of involving AFC memories
in terms of of the efficient entanglement distribution due to its multimodality and feasibility in the future quantum network.
\color{black}This paper is organized as follows. In \secref{cody}, we summarize the study from Ref.~\cite{Jones2016}. 
A protocol using the AFC is introduced in \secref{protocol}, the numerical simulation for the AFC protocol is shown in \secref{simu}, and the conclusions are presented in \secref{conclusion}.

\begin{figure}[t]

\includegraphics[scale=0.55]{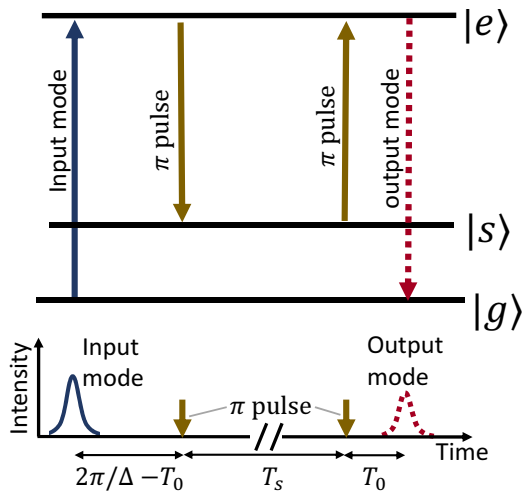}

\caption{\figlab{fig3}When a photon is absorbed by a comblike absorption profile, a transition from $\ket{g}$ to $\ket{e}$ occurs. When a $\pi$ pulse is applied, the population at the $\ket{e}$ level is transferred to $\ket{s}$.
Let $2\pi/\Delta - T_0$ be the time from the transition of the atom to $\ket{e}$ to the transition to $\ket{s}$. When a $\pi$ pulse is applied again, it transits from $\ket{s}$ to $\ket{e}$. Let $T_s$ be the time when the atom is at $\ket{s}$. After $T_0$, a transition from $\ket{e}$ to $\ket{g}$ occurs and a photon is emitted. }
\end{figure}

\begin{figure*}[t]
\mbox{
\includegraphics[scale=0.4]{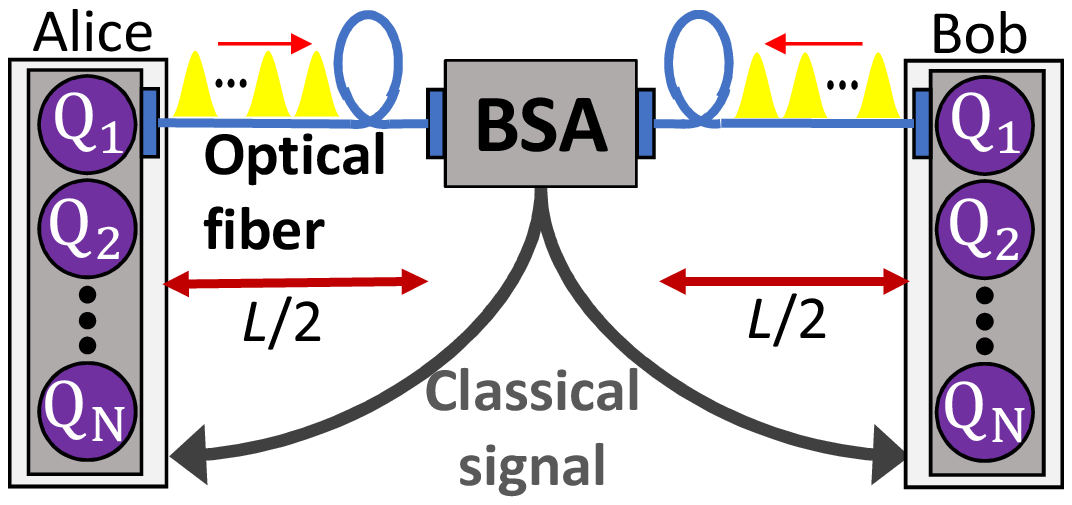}
{\small (a) }
} 
\mbox{
\includegraphics[scale=0.4]{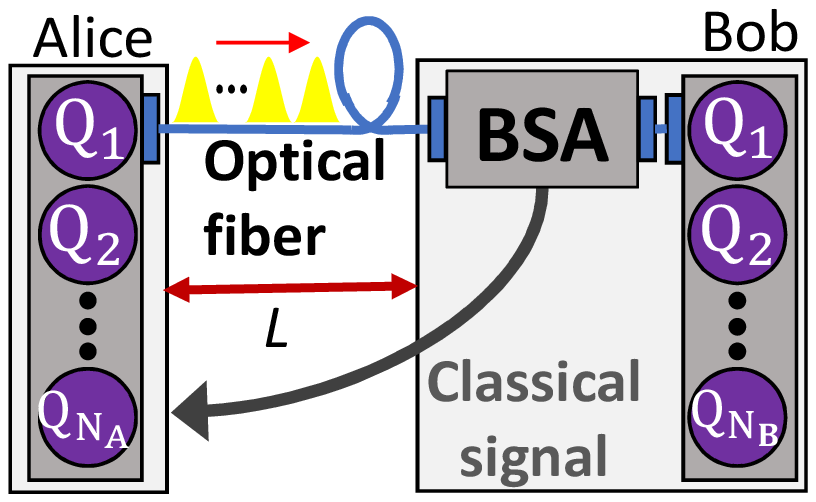}
{\small (b) }
} 
\mbox{
\includegraphics[scale=0.4]{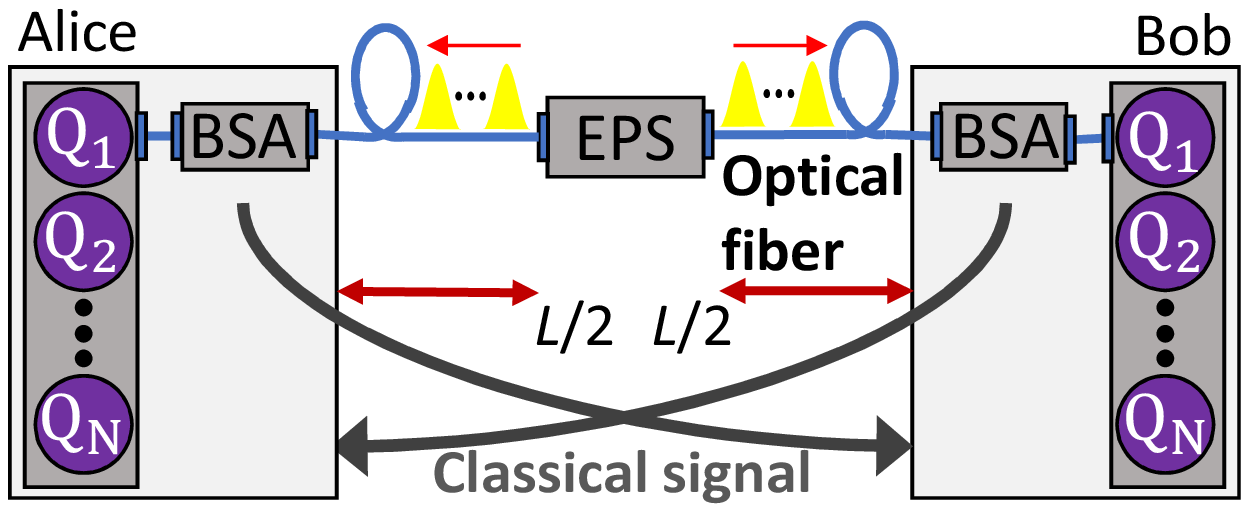}
{\small (c) }
}
\caption{\figlab{fig1}(a) Hardware arrangement for MM: the BSA is set in the middle of the two repeater nodes. Repeater nodes Alice and Bob simultaneously emit a photon which is entangled with a quantum memory $Q_i$. The result of the Bell measurement of whether entanglement sharing has been achieved is sent from the BSA to both Alice and Bob by classical communication. (b) Hardware arrangement for SR: the BSA is set inside Bob. (c) Hardware arrangement for MS: the EPS is at the intermediate point between Alice and Bob. Both Alice and Bob have a BSA inside each node and send the result of the Bell measurement to each other by classical communication.}
\end{figure*}

\section{Scheme of entanglement distribution}\seclab{cody}

In Ref.~\cite{Jones2016}, three types of schemes, MM, SR, and MS, were introduced and compared, which are illustrated in \figref{fig1}. In the following, we briefly outline the three schemes.

For MM, a photon entangled with a spin qubit inside a quantum memory is emitted from each side as in \figref{fig1}(a).
A Bell measurement of the two photons at the intermediate point entangles qubits in both memories. In this scheme, it is unknown whether the qubits in the two memories entangled until the outcome of the Bell measurement at the intermediate point is known via classical communication.

If the result of the Bell measurement from the Bell-state analyzer (BSA) immediately arrives at Alice and Bob, they can initialize their memories immediately when the outcome of the Bell measurement is false. 
Each memory can emit a photon entangled with a spin qubit once again without awaiting a classical communication for a long time. 
To minimize the time until they obtain the result of the Bell measurement, the BSA is placed inside Bob's node, as shown in \figref{fig1}(b) in the SR scheme.

In MS, Alice and Bob both become receivers by placing an EPS at the middle point, as the name suggests. Because BSAs are placed inside nodes, the success of the latest BSA simply indicates that a photon from the EPS arrives at one side, not that entanglement generation between the two nodes has been achieved. Entanglement generation is achieved by the simultaneous success of two Bell measurements.

We use the following assumptions and terminologies for the following simulation~\cite{Jones2016}.

\begin{itemize}
\item The number of quantum memories in each node in the MM and MS is $N$. In the case of SR, the number of memories on the sender (receiver) side is $N_{\rm A} (N_{\rm B})$, and $N_{\rm A} + N_{\rm B} = 2N$.
\item The time used for one trial is represented by $t_{\rm clock}$. $t_{\rm clock}$ is limited by the performance of the single-photon detector (SPD), EPS, and/or quantum memory. In addition, let $ t_{\rm link} = nL / c$ be the time required for photons in an optical fiber to travel between adjacent repeater nodes of distance $L$, where $n$ is the refractive index of the optical fiber and $c$ is the speed of light in vacuum. Here, the total synchronization time is set to $t_{\rm round}$.
\item The dark count rate shall be negligibly small. Therefore, with the efficiency of the SPD as $p_{\rm d}$, the success probability of the Bell measurement can be given by $p_{\rm BSA} = p_{\rm d} ^2/2$. Further, let $p_{\rm memory}$ be the probability that a photon is emitted from a memory and is coupled to an optical fiber. The optical fiber standard attenuation length is given by $L_{\rm att}$, and let $p_{\rm optical} = p_{\rm memory} \exp (-L / 2L_{\rm att})$ be the probability that a photon is emitted by a quantum memory and transmits through an optical fiber successfully. 
\item For MS, let the probability an entangled photon pair is generated at the EPS be $p_{\rm m}$. The probability $p_{\rm l} (p_{\rm r})$ that latches a qubit on the Alice (Bob) side after one clock cycle is $p_{\rm l} = p_{\rm r} = p_{\rm m} p_{\rm BSA} p_{\rm optical}$. Therefore, let $ K = \lceil N / p_{\rm m} p _{\rm BSA}p _{\rm optical}\rceil$ be the number of trials performed during the entire synchronization time $t_{\rm round}$.
\end{itemize}

\begin{table}[b]
\caption{
Performance of quantum memory. The values have been taken from Ref.~\cite{Jones2016}.
}\tablab{tab1}
\begin{ruledtabular}
\begin{tabular}{p{2cm} p{2cm} p{2cm} p{2cm}}
\textrm{Memory type}&
\textrm{Cycle time}&
\textrm{Emission fraction}&
\textrm{Collection efficiency}\\
\colrule
Trapped \\ ion ($^{\rm 171}\rm Yb^{\rm +}$)& 1~$\rm \mu$s & 1.00 & 0.05\\
Diamond NV & 100 ns & 0.50 & 0.50\\
Quantum \\ dot (InGaAs)& 10 ns & 0.90 & 0.50\\
\end{tabular}
\end{ruledtabular}
\end{table}

\begin{figure*}[t]
\mbox{
\includegraphics[scale=0.31]{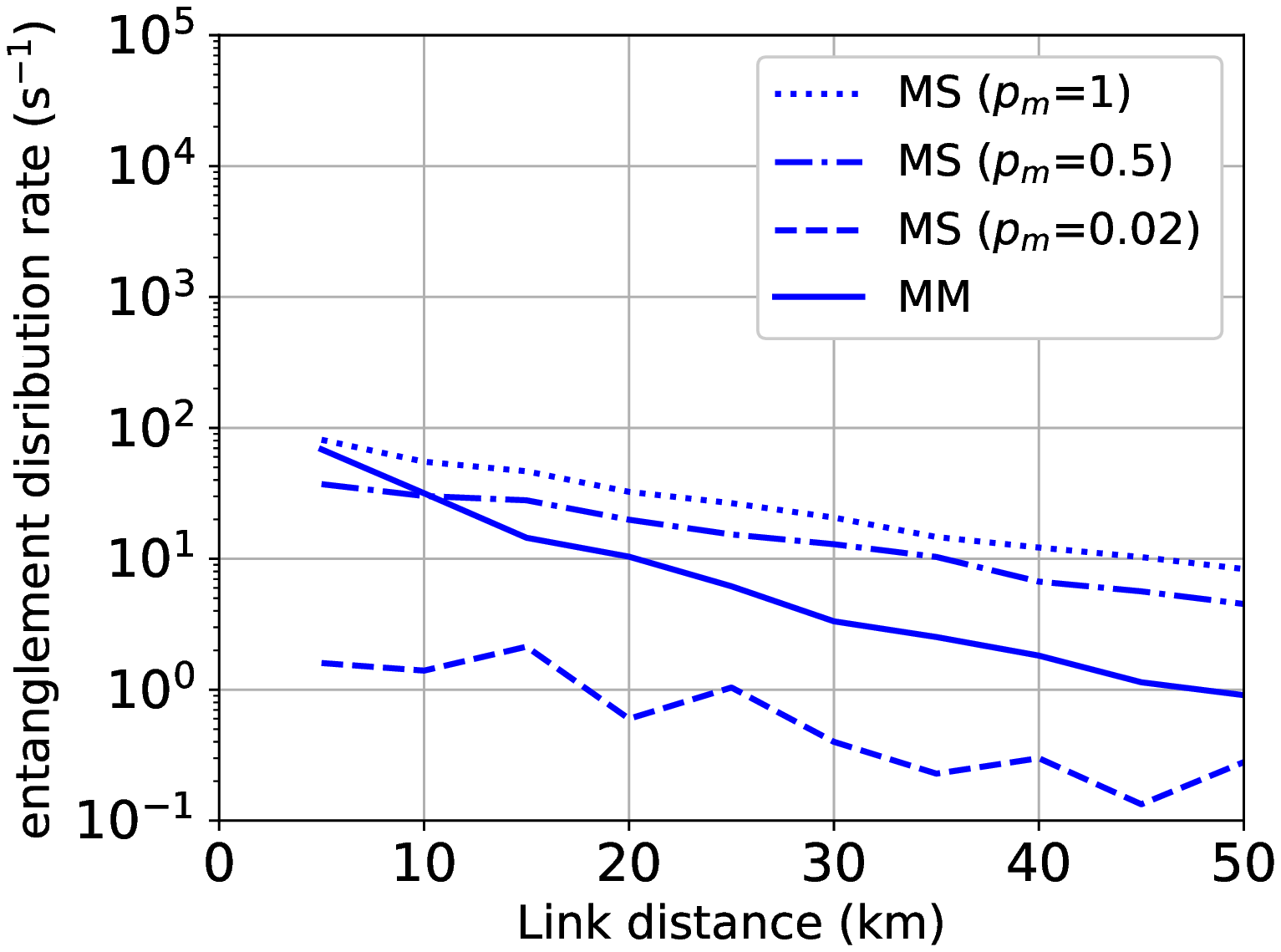}
{\small (a) }
}
\mbox{
\includegraphics[scale=0.31]{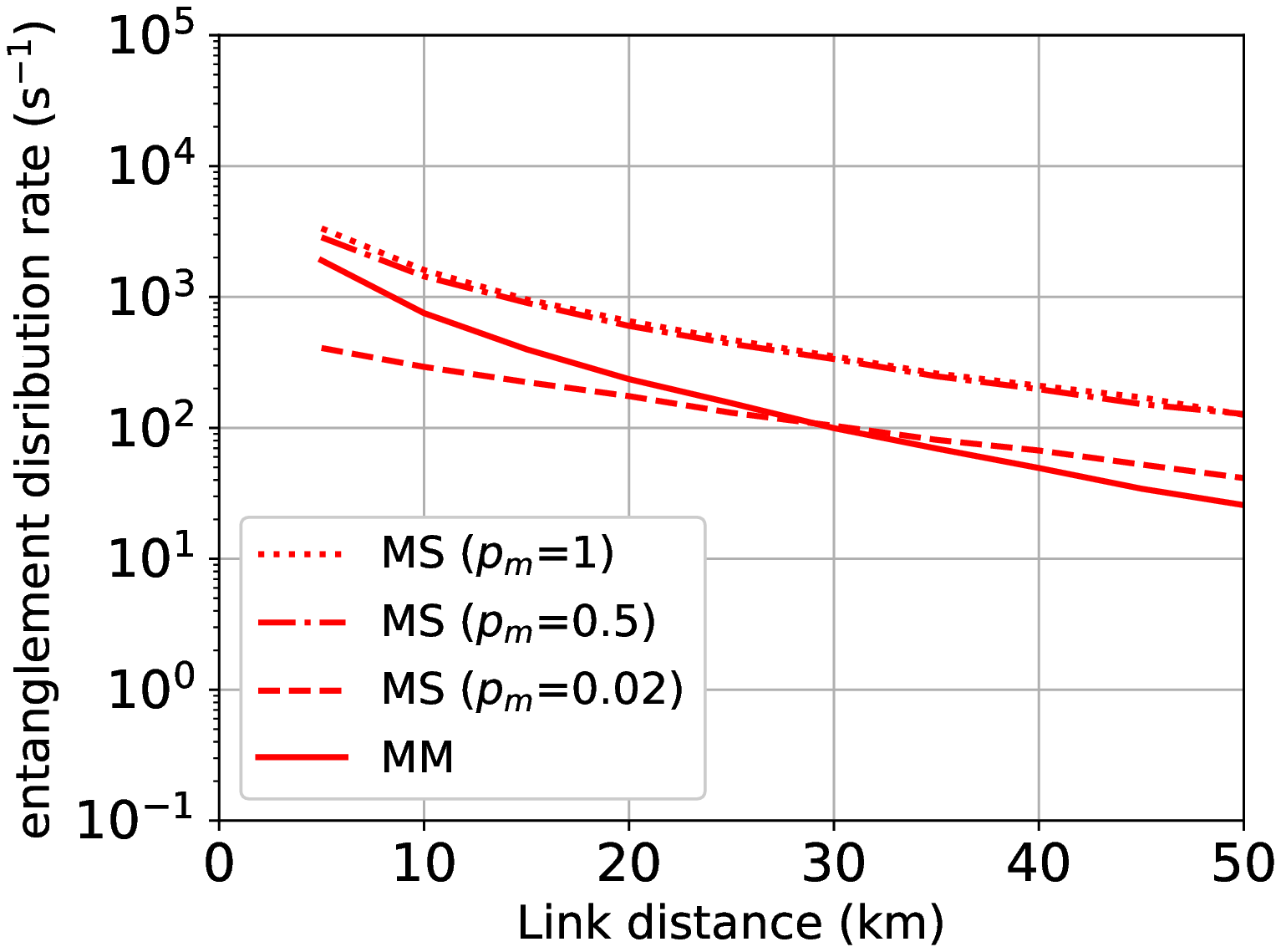}
{\small (b) }
}
\mbox{
\includegraphics[scale=0.31]{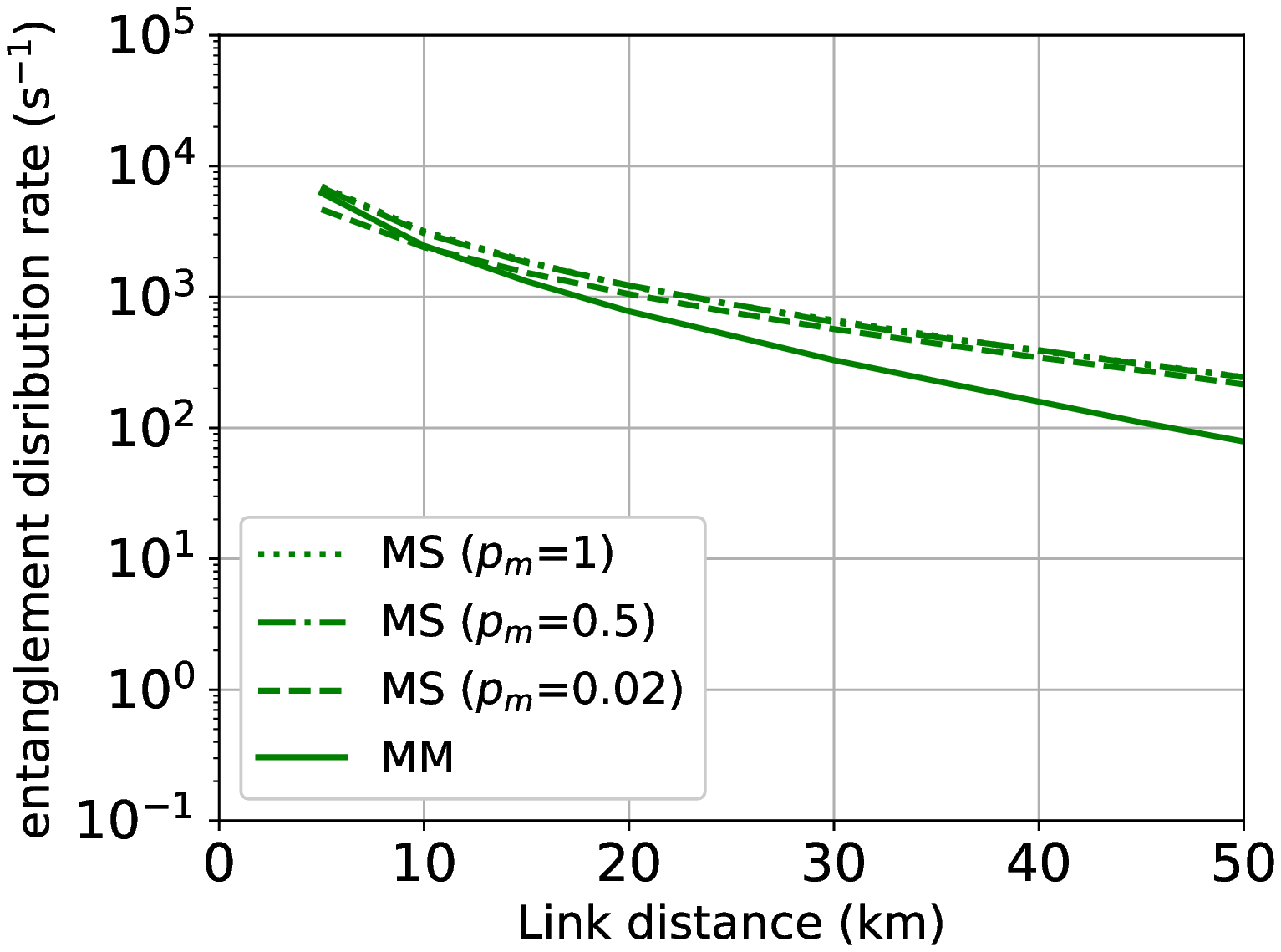}
{\small (c) }
}
\caption{\figlab{fig2} 
\color{black}Entanglement distribution rate of the scheme using spin-photon entanglement type quantum memories.($N = 3$) \color{black}
(a) Trapped ion: $p_{\rm optical} = 0.05 \exp (-L / 2L_{\rm att})$, $t_{\rm clock}=1\rm \mu$s; (b) diamond NV center: $p_{\rm optical} = 0.25 \exp (-L / 2L_{\rm att})$, $t_{\rm clock}=100$ns; (c) QD: $p_{\rm optical} = 0.45\exp (-L / 2L_{\rm att}), t_{\rm clock}=10$ns.}
\end{figure*}

Here, we compare the performances of each protocol.
Let $p ^{\rm MM}, p ^{\rm SR}, p ^{\rm MS}$, $t_{\rm round} ^{\rm MM}, t_{\rm round}^{\rm SR}, t_{\rm round} ^{\rm MS}$, and $R^{\rm MM}, R^{\rm SR}, R^{\rm MS}$ be the probabilities that an entanglement can be generated in the first trial, the total synchronization times, and the entanglement distribution rates of MM, SR, and MS, respectively.
Here, the parameters are given by

\begin{eqnarray}
p^{\rm MM} &=& p_{\rm BSA}(p_{\rm optical})^2,\equlab{eq1}\\
p^{\rm SR} &=& p_{\rm BSA}(p_{\rm optical})^2,\equlab{eq2}\\
p^{\rm MS} &=& p_{\rm m}(p_{\rm BSA}p_{\rm optical})^2,\equlab{eq3}
\end{eqnarray}

\begin{eqnarray}
t_{\rm round}^{\rm MM} &=& t_{\rm link}+Nt_{\rm clock},\equlab{eq4}\\
t_{\rm round}^{\rm SR} &=& 2t_{\rm link}+N_{\rm A}t_{\rm clock},\equlab{eq5}\\
t_{\rm round}^{\rm MS} &=& t_{\rm link}+Kt_{\rm clock},\equlab{eq6}
\end{eqnarray}

and, assuming that the approximation of $Kt_{\rm clock} \ll t_{\rm link} $ holds~\cite{Jones2016},

\begin{eqnarray}
R^{\rm MM} &=& \frac{Np_{\rm BSA}(p_{\rm optical})^2}{t_{\rm link}},\equlab{eq7}\\
R^{\rm SR} &<& \frac{N_{\rm A}p_{\rm BSA}(p_{\rm optical})^2}{2t_{\rm link}} < \frac{Np_{\rm BSA}(p_{\rm optical})^2}{t_{\rm link}},\equlab{eq8}\\
R^{\rm MS} &\approx& \frac{Np_{\rm BSA}p_{\rm optical}}{2t_{\rm link}}.\equlab{eq9}
\end{eqnarray}

At this point, it can be seen from \equref{eq7} and \equref{eq8} that the rate of SR is always lower than that of MM \color{black}(on condition that $N_{\rm A} +N_{\rm B} = 2N$)\color{black}. From \equref{eq7} and \equref{eq9}, the ratio between them can be given by
\begin{equation}
\frac{R^{\rm MS}}{R^{\rm MM}} \approx \frac{1}{2p_{\rm optical}} = \frac{1}{2p_{\rm memory}\exp (-\frac{L}{2L_{\rm att}})}. \equlab{eq10}
\end{equation}

Therefore, their relative performance depends on the efficiency of the spin-photon entanglement generation $p_{\rm memory}$ and the distance $L$ between adjacent nodes. Note that the assumption of $Kt_{\rm clock} \ll t_{\rm link} $ becomes invalid when $p_{\rm m}$ becomes small. Here, we focus on a situation where a relatively large $p_{\rm m}$ is available.

Here, \color{black}we attempt to reproduce the result of Ref.~\cite{Jones2016}. \color{black}
The simulation by the Monte Carlo method is carried out with MM and MS for the listed memories in \tabref{tab1} using the realistic parameters taken from Ref.~\cite{Jones2016}. 
The simulation results are presented in \figref{fig2}.
The plots in \figref{fig2} are the averages of $10^5t_{\rm round}$. Three values of 1, 0.5, and 0.02 were used for $p_{\rm m}$, and the number of memories $N$ was set to $N = 3$~\cite{Jones2016}. 
In addition, $L_{\rm att} = 22$ km, $c = 2.998 \times 10^5 $ km/s, $n = 1.5$, and $p_{\rm d} = 0.8$ were used. 
A quantum purification procedure for increasing fidelity was not included, and the dark count probability was ignored. 
The above analysis and the results of the numerical simulation agree with each other. 
The reason some differences from the results presented in Ref.~\cite{Jones2016} exist 
\color{black}seems to be \color{black}because of the difference in convergence caused by the different simulation methods. 
As the entanglement distribution rate decreases, 
the difference becomes noticeable.

\begin{figure*}[t]
\begin{minipage}{0.50\hsize}
\includegraphics[scale=0.5, clip, keepaspectratio]{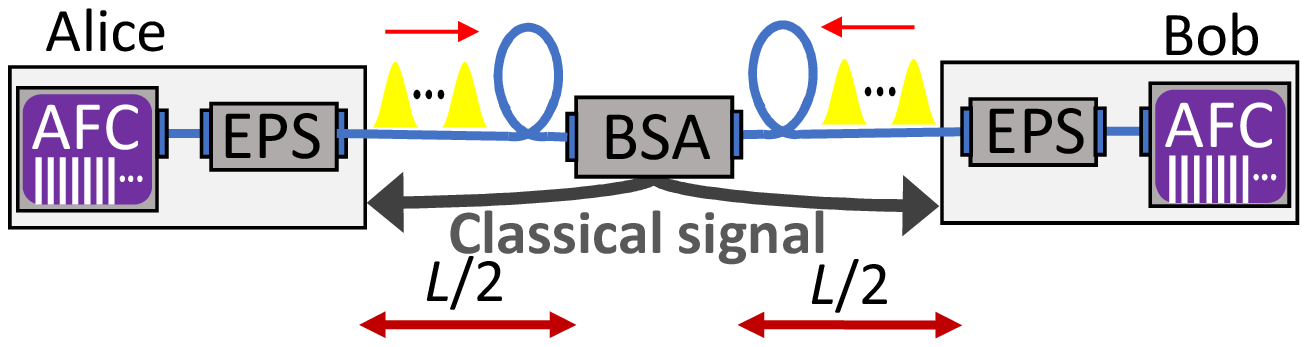}
{\small (a) }
\end{minipage}
\begin{minipage}{0.45\hsize}
\includegraphics[scale=0.5, clip, keepaspectratio]{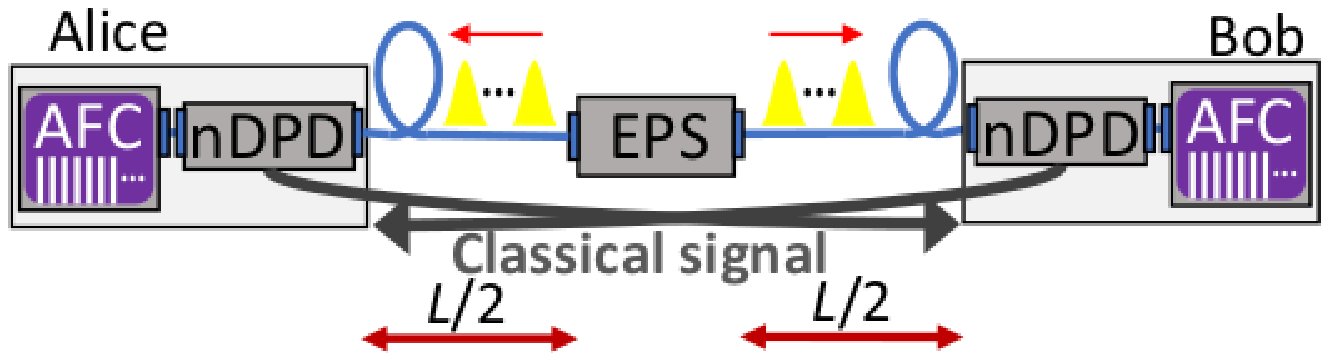}
{\small (b) }
\end{minipage} \\
\begin{minipage}{0.06\hsize}
\vspace{5mm}
\end{minipage} \\
\begin{minipage}{0.45\hsize}
\includegraphics[scale=0.4, clip, keepaspectratio]{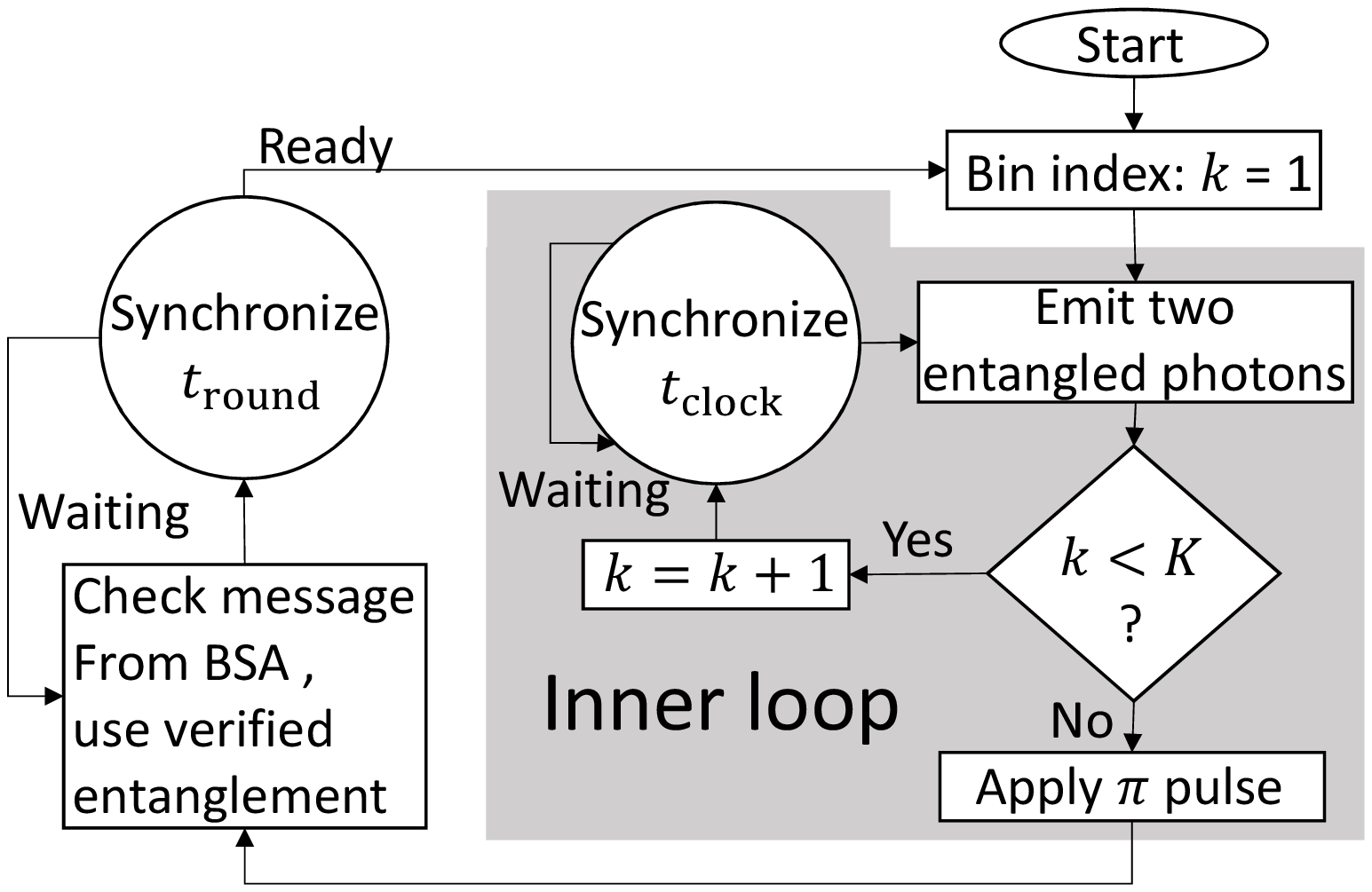}
{\small (c) }
\end{minipage} 
\begin{minipage}{0.45\hsize}
\includegraphics[scale=0.4, clip, keepaspectratio]{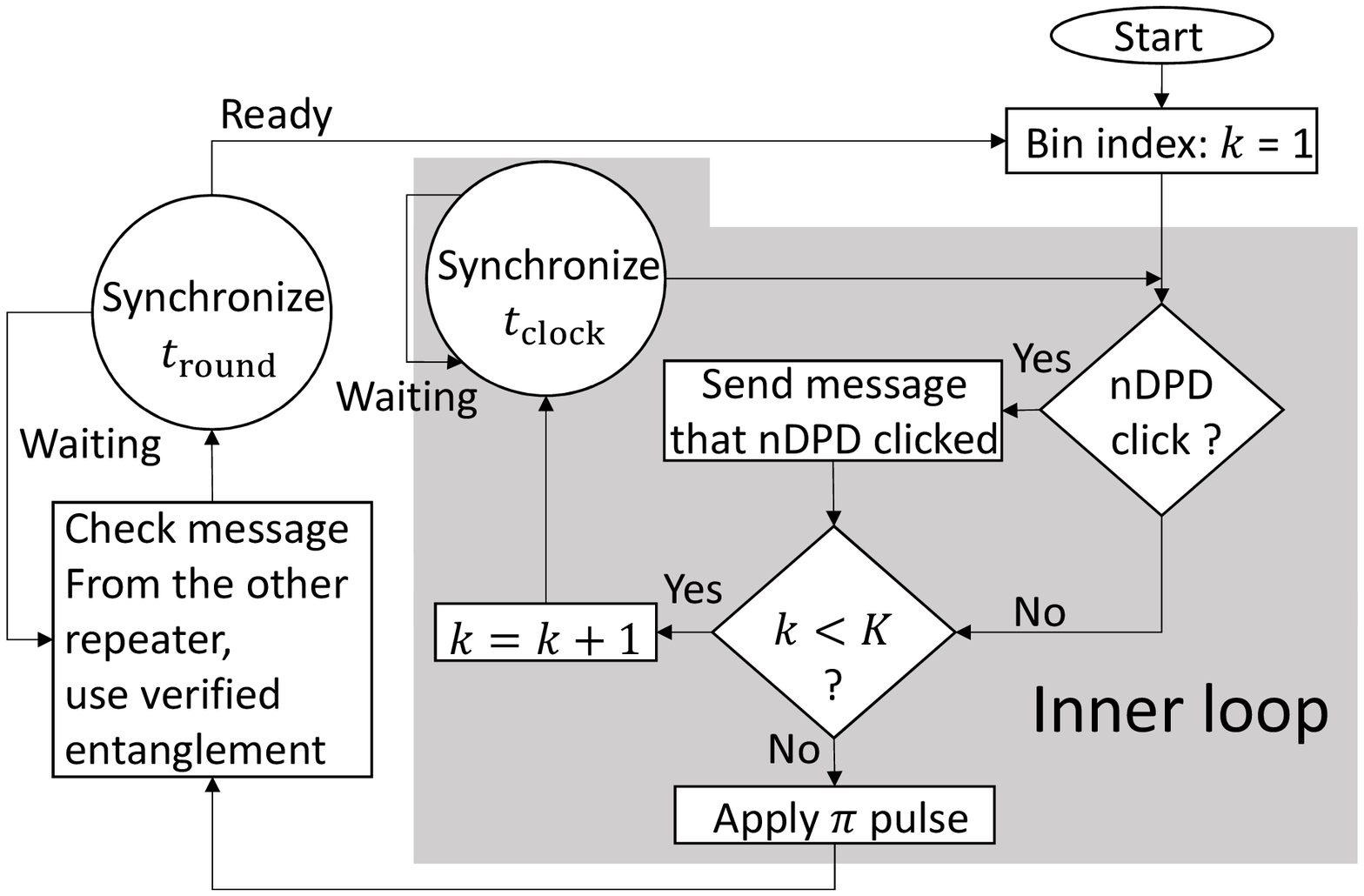}
{\small (d) }
\end{minipage} 
\caption{\figlab{fig4}(a) Schematic of AFC-MM: the BSA is set at the middle point between Alice and Bob. Each has an EPS and sends a photon from one entangled photon pair to the BSA simultaneously. (b) Schematic of AFC-MS: each node has an nDPD. When a photon arrives, the notification is sent to the other node. Flowchart for each receiver in (c) AFC-MM and (d) AFC-MS.}
\end{figure*}


\section{Protocol of AFC}\seclab{protocol}

\figref{fig4} presents the arrangements of hardware components and flowcharts for MM and MS for the adoption of AFC quantum memory. In this study, AFC quantum memory does not function as a spin-photon entanglement source but as an ‘absorbing’ quantum memory. Here, we do not consider SR using the AFC, as the entanglement distribution rate is clearly smaller than that of MM.

In the AFC-MM, the EPS is arranged in the immediate vicinity of the memory in the repeater node. Here, the loss of photons from the EPS to memory is ignored.
In the AFC-MS, we assume non-destructive photon detectors (nDPDs). By placing an nDPD before the AFC, the nDPD click functions as a weak heralding signal.
\color{black}It is great advantage because the scheme is not affected by inefficiency of Bell state analyzer. \color{black}
Of course, it is difficult to implement the nDPD itself.
As an nDPD, the AFC will also be one of the promising candidates. In Ref.~\cite{Sinclair2016}, the \color{black}proof-of-principal of \color{black}non-destructive detection of photonic qubits using the phase shift by the a.c. Stark shift was demonstrated.

With the above scenarios for the AFC-MM and AFC-MS, it is not guaranteed that a photon will be reliably absorbed but does confirm the presence of a photon immediately before the AFC.
Note that photon loss causes this imperfection. Therefore, it affects efficiency of entanglement swapping between adjacent links, but not the fidelity of quantum state.
However, even in this ``weak’’ heralding, the quality of heralding improves, for example, by (even probabilistically) the detection of photons not absorbed by the AFC memory. It will be very practical if an AFC absorption efficiency close to 100\% is realized in the future; however, an inefficient heralding scheme can still be useful for entanglement distribution. In the present simulation, we will only discuss the entanglement distribution rate between adjacent repeater nodes.
Also same as Ref.~\cite{Jones2016}, we assume that the dark counts of the detector and multiple photon pair emissions at EPS, which affect the fidelity of the entangled state, are small enough to be suppressed by a purification procedure.
Miliherz dark count rates~\cite{Schuck2013} and a high performance entangled-photon source~\cite{PhysRevLett.122.113602} have already been achieved.

Let the absorption efficiency of the AFC be $p_{\rm AFC}$, and let $p_{\rm optical}^{\prime} = p_{\rm AFC} \exp (- L / 2 L_{\rm att})$. With the AFC-MM, the probability $p^{\rm MM ^{\prime}}$ that entanglement can be shared in a single trial can be given by

\begin{equation}
\begin{split}
p^{\rm MM^{\prime}} &= p_{\rm BSA}(p_{\rm m}p^{\prime}_{\rm optical})^2\\
&=p_{\rm BSA}(p_{\rm m}p_{\rm AFC} \exp(-\frac{L}{2L_{\rm att}}))^2 . \equlab{eq11}
\end{split}
\end{equation}

In addition, when the efficiency of the nDPD is given by $p_{\rm pass}$ in the AFC-MS, the probability $p^{\rm MS ^{\prime}}$ that entanglement can be shared in a single trial can be given by

\begin{equation}
\begin{split}
p^{\rm MS ^{\prime}} &= p_{\rm m}(p_{\rm pass}p^{\prime}_{\rm optical})^2\\
&= p_{\rm m}(p_{\rm pass}p_{\rm AFC}\exp(-\frac{L}{2L_{\rm att}}))^2.\equlab{eq12}
\end{split}
\end{equation}

Here, the probability 
$p_{\rm X}^{\rm Y}$ ($\rm X =$ Alice or Bob, $\rm Y =\rm MM^{\prime}$ or $\rm MS^{\prime}$) 
that can latch a qubit on the X side at Y in one clock cycle is 
\begin{eqnarray}
p_{\rm X}^{\rm MM^{\prime}} = p_{\rm m}p_{\rm AFC},\\ \equlab{addeq2}
p_{\rm X}^{\rm MS^{\prime}} = p_{\rm m}p_{\rm pass}p_{\rm optical}^{\prime}. \equlab{addeq3}
\end{eqnarray}
Therefore, when the number of modes is $N_{\rm AFC}$,
the number of trials $K ^{\rm Y}$ can be basically given by
\begin{equation}
K ^{\rm Y} = \lceil N_{\rm AFC} / p_{\rm X}^{\rm Y}\rceil . \equlab{addeq4}
\end{equation}
With $ t_{\rm clock}^{\prime}$ as the time used for one trial at Y,
for the case the parameters fulfill the following relation, 
$K^{\rm Y}t_{\rm clock}^{\prime} > 2\pi / \Delta$, 
the first absorbed photon will be re-emitted by rephasing. Therefore, in the conditions, they are given as
\begin{equation}
K^{\rm Y} = \lceil (2\pi / \Delta) / t_{\rm clock}^{\prime}\rceil . \equlab{addeq5}
\end{equation}
Assuming that
$K ^ {\rm Y} = \lceil N_{\rm AFC}/p_{\rm X}^{\rm Y} \rceil$,
and 
$K ^ {\rm Y}t_{\rm clock}^{\prime} \ll t_{\rm link}$, 
the rate of the AFC-MM
$R^{\rm MM^{\prime}}$
and rate of the AFC-MS
$R^{\rm MS^{\prime}}$
can be given by

\begin{equation}
\begin{split}
R^{\rm MM^{\prime}} &= \frac{K ^{\rm MM^{\prime}}p^{\rm MM^{\prime}}}{t_{\rm link}}\\
&\simeq \frac{N_{\rm AFC}p_{\rm BSA}p_{\rm m}p_{\rm AFC}(\exp(-\frac{L}{2L_{\rm att}}))^2}{t_{\rm link}}\equlab{eq13},
\end{split}
\end{equation}

\begin{equation}
R^{\rm MS^{\prime}} \approx \frac{N_{\rm AFC}p_{\rm pass}p_{\rm AFC}\exp(-\frac{L}{2L_{\rm att}})}{2t_{\rm link}}\equlab{eq14}.
\end{equation}

For \equref{eq14}, an approximation similar to \equref{eq9} was made.
The ratio between the AFC-MM and AFC-MS is as follows:

\begin{equation}
\frac{R^{\rm MS^{\prime}}}{R^{\rm MM^{\prime}}} = \frac{p_{\rm pass}}{2p_{\rm BSA}p_{\rm m}\exp(-\frac{L}{2L_{\rm att}})}\equlab{eq15}.
\end{equation}

From \equref{eq15}, as the transmission distance increases, the rate of MS also relatively increases.
Next, the ratio between the rates of the AFC-MS, AFC-MM, and MS can be given by

\begin{equation}
\frac{R^{\rm MM^{\prime}}}{R^{\rm MS}} = \frac{2N_{\rm AFC}p_{\rm m}p_{\rm AFC}\exp(-\frac{L}{2L_{\rm att}})}{Np_{\rm memory}},\equlab{eq16}
\end{equation}

\begin{equation}
\frac{R^{\rm MS^{\prime}}}{R^{\rm MS}} = \frac{N_{\rm AFC}p_{\rm AFC}p_{\rm pass}}{Np_{\rm BSA}p_{\rm memory}}.\equlab{eq17}
\end{equation}

It is still difficult to multiplex quantum memories such as QDs, NV centers in diamond, and trapped ions; therefore, $N$ is realistically limited to one digit, as assumed to be three in Ref.~\cite{Jones2016}, whereas $N_{\rm AFC}$ can be more than $10^2$~\cite{Jobez2016}. 
Within a certain distance range, the AFC-MM exceeds the rate of the MS in \secref{cody}.
In addition, if the following relations are valid: 
$p_{\rm AFC} \approx p_{\rm memory}$ and 
$p_{\rm pass} \approx p_{\rm BSA}$ for \equref{eq17}, which appears to be realistic assumptions, the entanglement distribution rate of the AFC-MS will exceed the rate of MS at \secref{cody} considerably from the fact of $N_{\rm AFC} \gg N$.

\begin{figure*}[t]
\mbox{
\includegraphics[scale=0.4]{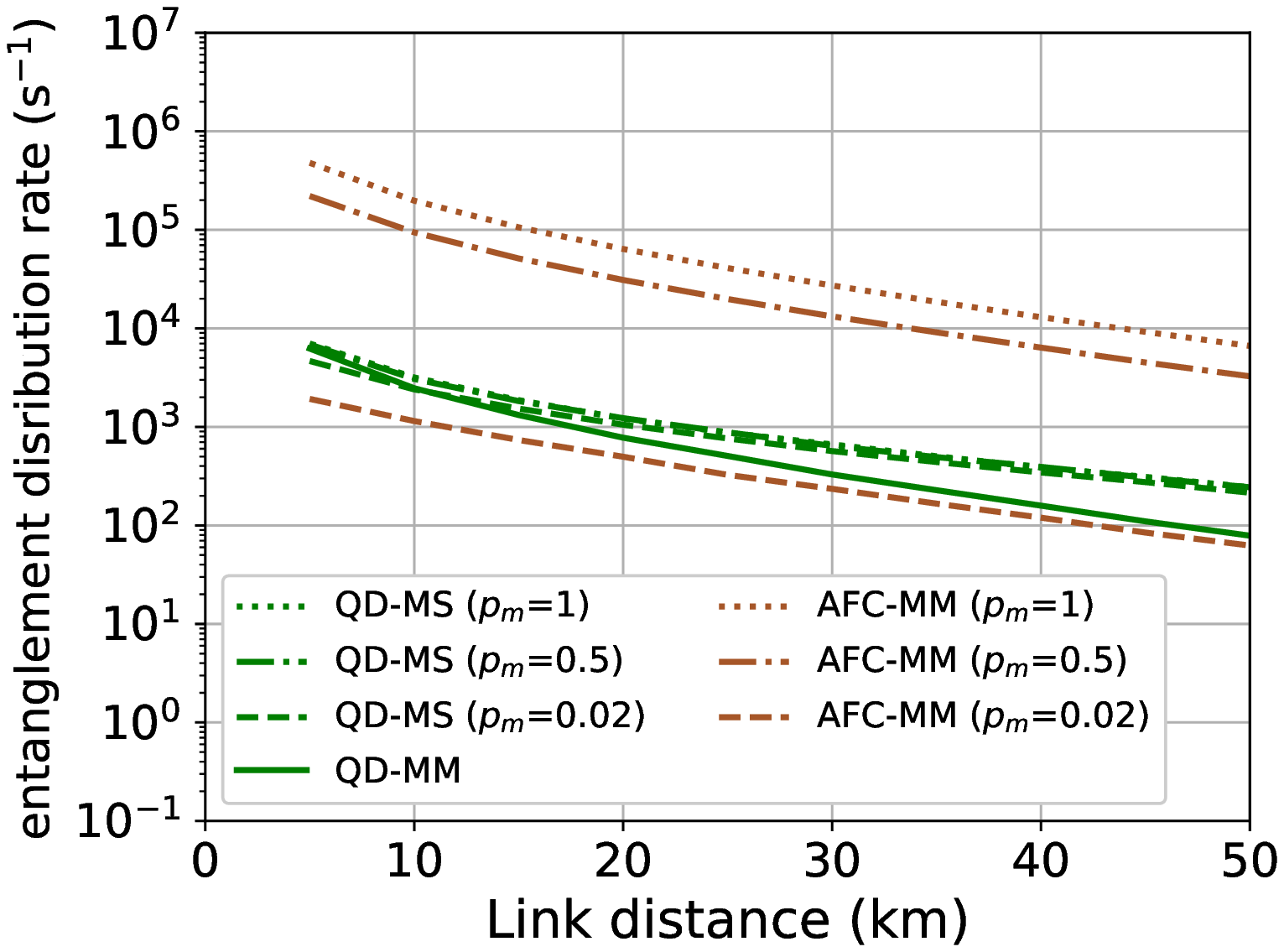}
{\small (a) }
} 
\mbox{
\includegraphics[scale=0.4]{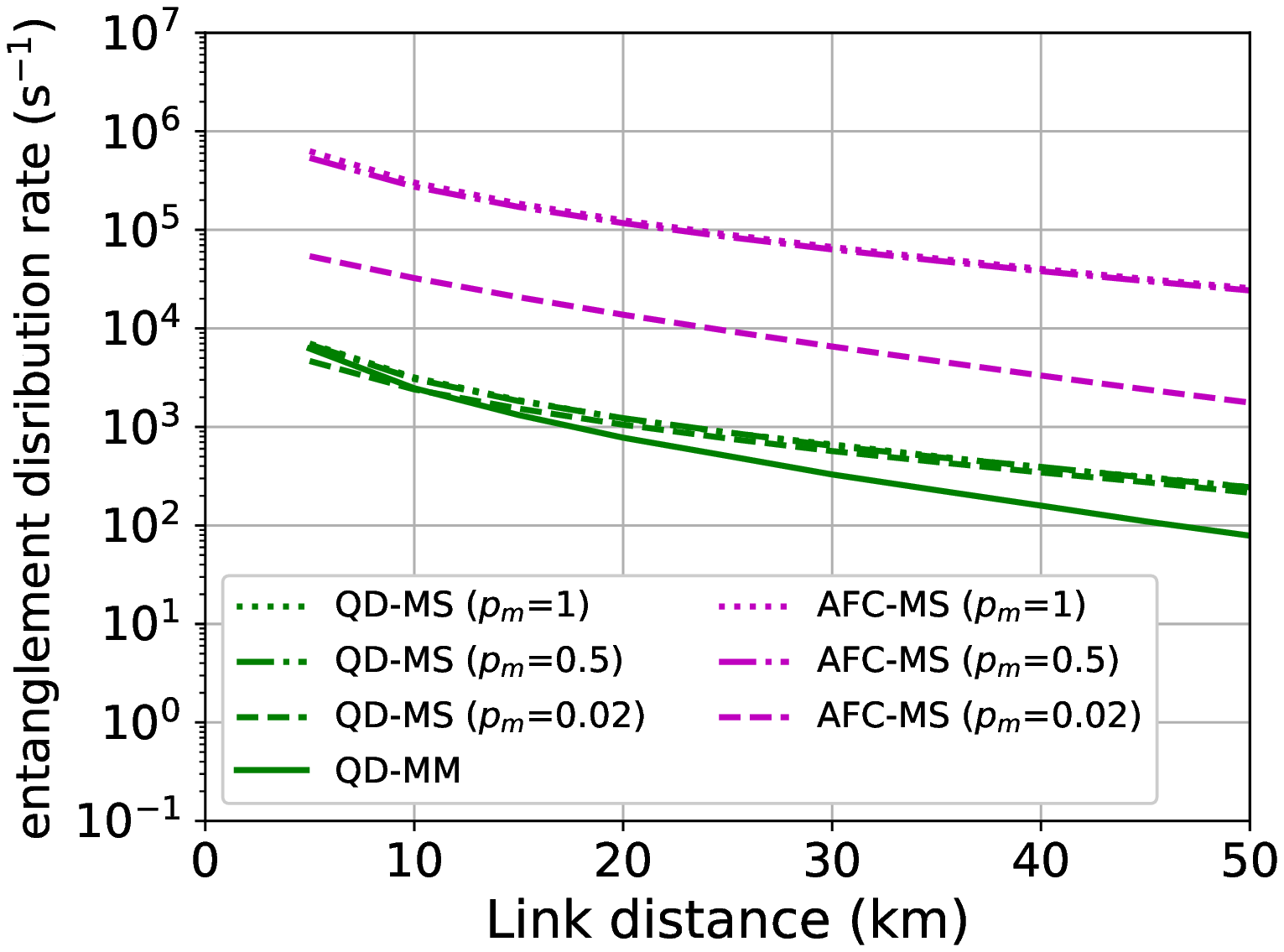}
{\small (b) }
} 
\mbox{
\includegraphics[scale=0.4]{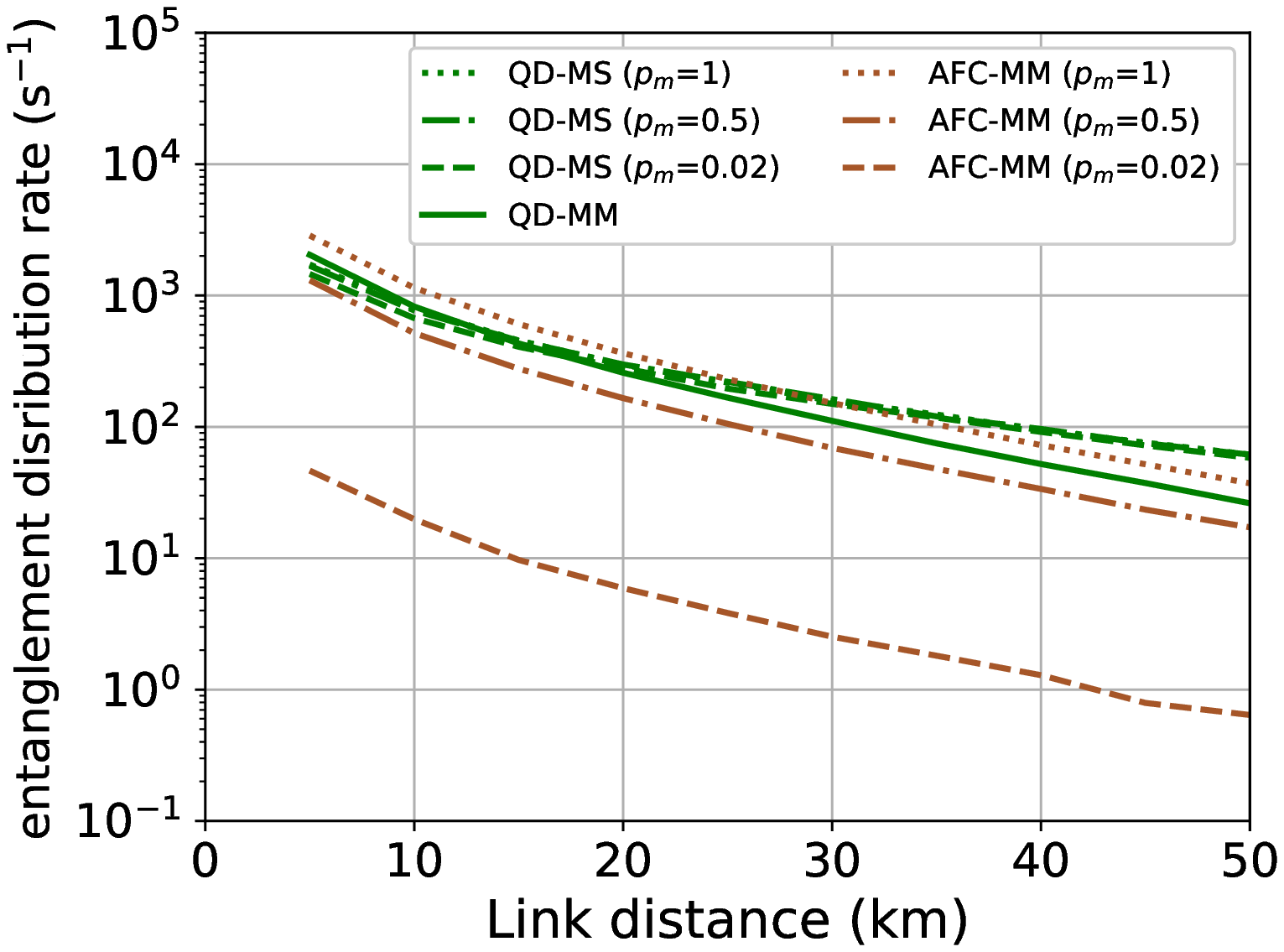}
{\small (c) }
} 
\mbox{
\includegraphics[scale=0.4]{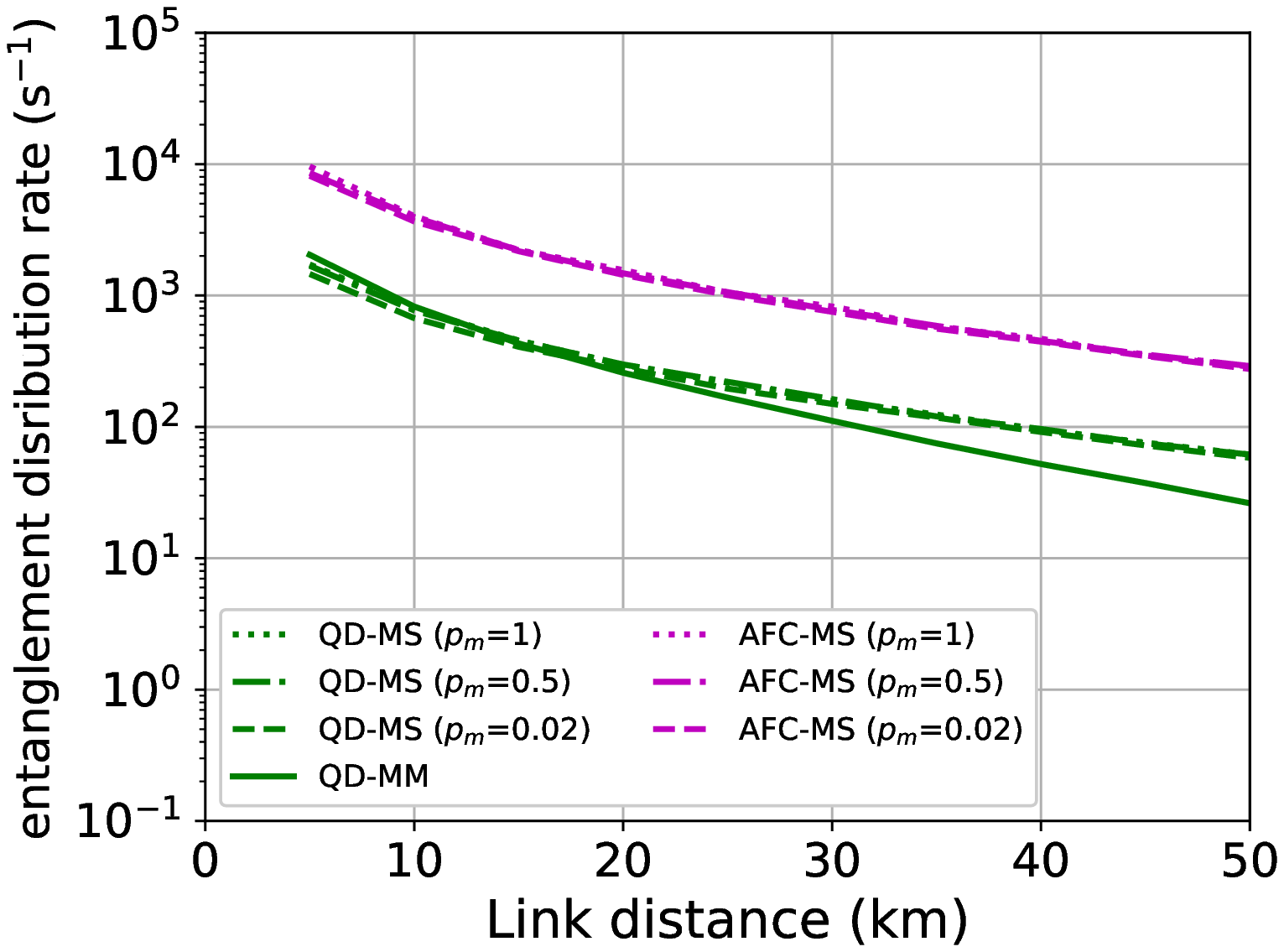}
{\small (d) }
} 
\caption{\figlab{fig5}Entanglement distribution rate of AFC-MM and AFC-MS. The green lines \color{black}in (a) and (b) \color{black} are the same as \figref{fig2}(c) \color{black}and the green lines in (c) and (d) are the results of simulations using QD in the case of $N=1$. \color{black}(a)AFC-MM\color{black}($N_{\rm AFC}=100$)\color{black} : $p_{\rm optical} = 0.53 \exp (-L / 2L_{\rm att})$, $t_{\rm clock}=10$ ns;
(b)AFC-MS\color{black}($N_{\rm AFC}=100$)\color{black} : $p_{\rm optical} = 0.53 \exp (-L / 2L_{\rm att})$, $t_{\rm clock}=10$ ns; 
\color{black}(c)AFC-MM($N_{\rm AFC}=1$) : $p_{\rm optical} = 0.53 \exp (-L / 2L_{\rm att})$, $t_{\rm clock}=10$ ns; 
(d)AFC-MS($N_{\rm AFC}=1$) : $p_{\rm optical} = 0.53 \exp (-L / 2L_{\rm att})$, $t_{\rm clock}=10$ ns.\color{black}}
\end{figure*}

\begin{figure*}[t]
\mbox{
\includegraphics[scale=0.4]{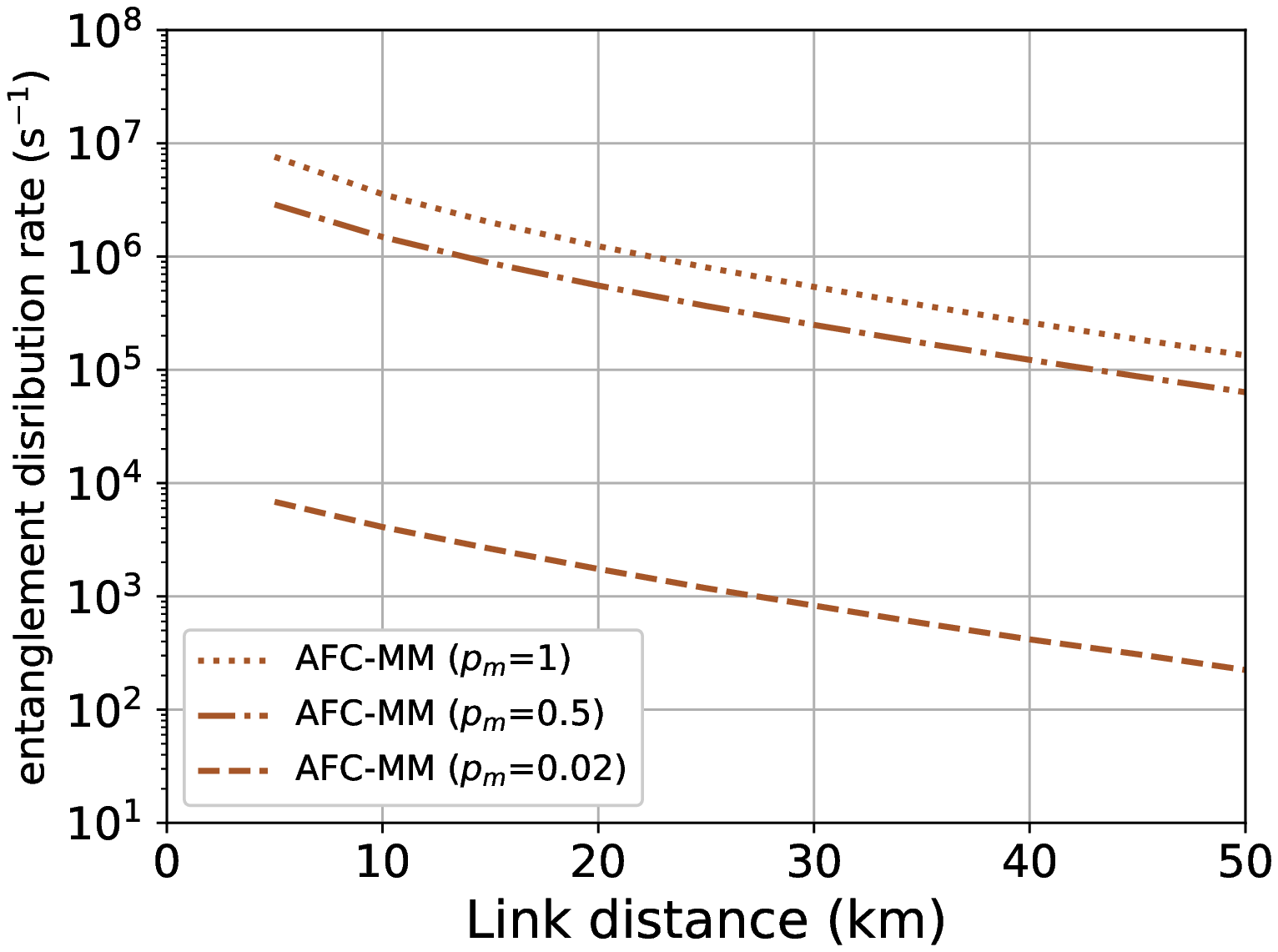}
{\small (a) }
} 
\mbox{
\includegraphics[scale=0.4]{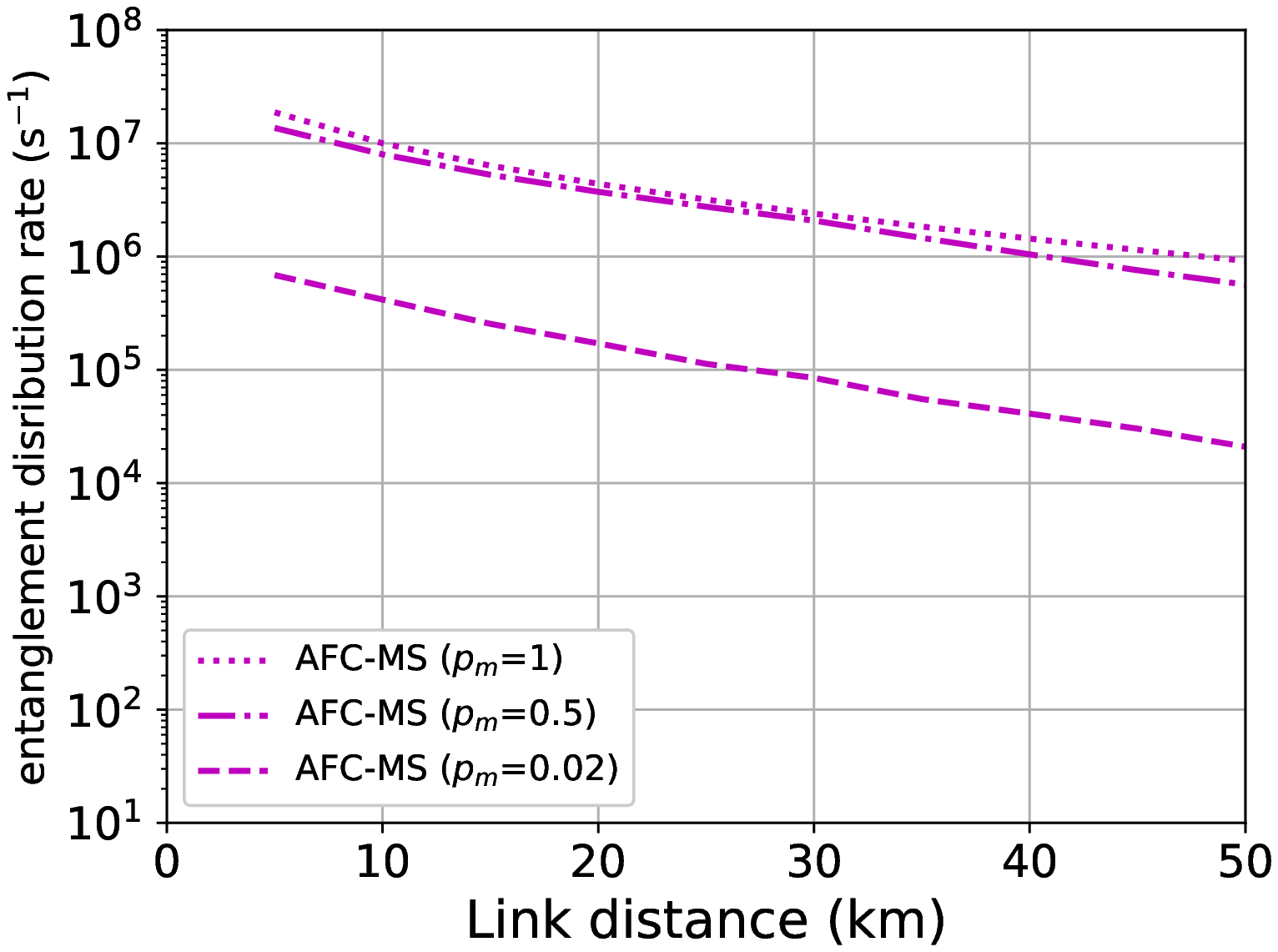}
{\small (b) }
} 
\caption{\figlab{fig6}Optimistic entanglement distribution rate of optimistic AFC-MM and optimistic AFC-MS. (a)AFC-MM: $p_{\rm optical} = \exp (-L / 2L_{\rm att})$, $t_{\rm clock}=10$ ns; (b)AFC-MS: $p_{\rm optical} = \exp (-L / 2L_{\rm att})$, $t_{\rm clock}=10$ ns. }
\end{figure*}

\section{Simulation}\seclab{simu}

From recent studies on AFCs using Eu:YSO as a REIDS, we perform an optimistic simulation using reasonable values for each parameter. The following assumptions and terminologies for the simulation are used:

\begin{itemize}
\item In \secref{cody}, as 
$t_{\rm clock}$ for the QD was set to 10 ns, the EPS and BSA have a standby time within 10 ns. We use the same value for the present study $t^{\prime}_{\rm clock} = 10$ ns.
\item The AFC's
\color{black}rephasing \color{black} time is $2\pi/\Delta =51\rm\mu$s~\cite{Jobez2016},
\color{black}
and coherent time at $\ket{s}$ is $1$ ms \cite{Jobez2015}.
\color{black}
The maximum time required for 1 round is 
$2\pi/\Delta + nL_{\rm max}/c \simeq 51\rm\mu$s $ + 250\rm\mu$s $= 301\rm\mu$s in our scheme.
Therefore, 1 ms, which is coherent time, \color{black}is sufficient for quantum repeater \color{black}in our scheme\color{black}.
\color{black}
\item $N_{\rm AFC} = 100$~\cite{Jobez2016} was used.
\item In Ref.~\cite{Jobez2014}, as the total efficiency of absorption and emission, 53\% was achieved. Assuming that the absorption probability of the AFC is pessimistically equivalent to the total efficiency, 
$p_{\rm AFC} = 0.53$.
\item \color{black}According to supplementary information of Ref.~\cite{Sinclair2016}, the signal loss due to nDPD less than 0.1 can be achieved by using a cavity. Therefore, we used $p_{\rm pass}= 0.9$ optimally.
\color{black}

\item $L_{\rm att} = 22$ km, 
$c = 2.998 \times 10^5$ km/s, $n = 1.5$, and $p_{\rm d} = 0.8$ were used.

\item The efficiency of the entangled two-photon source is simulated with three values of $p_{\rm m} = 1, 0.5$, and $0.02$, respectively~\cite{Jones2016}.
\item Calculate the average of $5 \times10^5 t_{\rm round}$ \color{black}by the Monte Carlo method within the range not exceeding the memory capacity in one round \color{black}
for each $5$ km from $L = 5$ to $50$ km~\cite{Jones2016}.
\end{itemize}

The results are presented in \figref{fig5}(a) and (b). 
\color{black}To compare the schemes themselves, the results of simulations assuming that $N_{\rm AFC} = N = 1$ are also shown in \figref{fig5}(c) and (d).
As the simulation in \secref{cody}, a purification procedure is not included, and the dark count is ignored.
\color{black} We are mainly concerned for photon loss.

Comparing \figref{fig5}(a) to the rate of \figref{fig5}(b), the rate of the AFC-MS is generally higher and the ratio, $R^{\rm MS^{\prime}}/R^{\rm MM^{\prime}}$, increases with the increase in distance $L$ and/or decrease in $p_{\rm m}$. Comparing \figref{fig5}(b) to \figref{fig2}(c), when $p_{\rm m} = 1, 0.5$, the rate of the AFC-MS exceeds that of the QD-MM by nearly two orders of magnitude, where the difference in mode number \color{black}and whether BSAs are used or not are \color{black} 
responsible. In the AFC using Tm:YAG as the REIDS, $N_{\rm AFC} = 1060$~\cite{Bonarota2011} was achieved. Therefore, a further rate improvement can be expected.

Therefore, a simulation, using the number of modes $N_{\rm AFC} = 1060$~\cite{Bonarota2011}, $2\pi / \Delta = \color{black}51\rm \mu\color{black}$\color{black}s~\cite{Jobez2016}, coherent time at $\ket{s}$ is $1$ ms \cite{Jobez2015} \color{black} as the optimal value, and absorption efficiency $p_{\rm AFC}$ = 1, is performed and presented in \figref{fig6}. If $p_{\rm m}$ is not extremely small, an entanglement distribution rate of approximately $1$ MHz at $L = 50$ km can be achieved.

Issues in implementing the protocol using the AFC is in the process of entanglement swapping between nodes. Issues related to it in the AFC are resulted from the entanglement purification and heralding.
It is difficult to purify high-fidelity entanglement under the present technology. In addition, as mentioned in \secref{protocol}, the scheme proposed in the present study does not become a perfect heralding unless the absorption rate of the AFC is 100\%. 
Let us consider the effect of this imperfection on entanglement swapping using linear-optic Bell-state measurement between adjacent elementary links.
Here, we assume that $J$ entanglement photon pairs are shared respectively in both links in AFC-MS.
Let $K_{\rm swap}$, $p_{\rm swap}$, and $p_{\rm emit}$ be the number of trials to swap entanglement, the success probability  of entanglement swapping in single trial, and the re-emission probability of photon from AFC.
In the case of perfect heralding (which means $p_{\rm pass}=1, p_{\rm AFC}=1$),
\begin{eqnarray}
K_{\rm swap} &=& J,\\ \equlab{addeq6}
p_{\rm swap} &=& p_{\rm emit}^2p_{\rm BSA}. \equlab{addeq7}
\end{eqnarray}
Therefore, expected number of successes is $Jp_{\rm emit}^2p_{\rm BSA}$ after $K_{\rm swap}$ trials.
In the case of imperfect heralding,
\begin{eqnarray}
K_{\rm swap} &=& \frac{J}{p_{\rm pass}p_{\rm AFC}},\\ \equlab{addeq8}
p_{\rm swap} &=& (p_{\rm pass}p_{\rm AFC})^2p_{\rm emit}^2p_{\rm BSA}, \equlab{addeq9}
\end{eqnarray}
due to the effect of the transmittance of nDPD and AFC.
Therefore, expected number of successes is $Jp_{\rm pass}p_{\rm AFC}p_{\rm emit}^2p_{\rm BSA}$ after $K_{\rm swap}$ trials.
Hence, after swapping over $i$ links, the rate for the case of imperfect heralding can be expected to be about $(p_{\rm pass}p_{\rm AFC})^{i-1}$ times that for perfect heralding.
For example, when $i = 10$,  $(p_{\rm pass}p_{\rm AFC})^{i-1}\approx10^{-3}$ with $p_{\rm pass}=0.9$ and $p_{\rm AFC}=0.53$.

\section{Conclusions}\seclab{conclusion}

We analyzed and simulated entanglement distribution rates between adjacent quantum repeater nodes using 
\color{black}
one
\color{black}AFC-type quantum memory
\color{black}
inside each quantum repeater 
\color{black}and demonstrated that the rate can be higher than the cases utilizing other types of memories
\color{black}which directly generate spin-photon entanglement and suffer from a difficulty of efficient generation and multimodality.
\color{black}The extension of the present scheme to a large number of repeater nodes for finding the best performance architecture remains for future studies, and the research of the AFC quantum gate, especially for implementing efficient entanglement purification, also remains for future studies.
When considering the result in this paper, AFC
can be the most preferable memory for high-rate entanglement distribution from the viewpoint of practical use.

\begin{acknowledgments}
This work was supported by 
the Toray Science foundation, 
the Asahi Glass foundation, 
KDDI foundation, 
SECOM foundation,
JST PRESTO JPMJPR1769, 
JST START ST292008BN, 
and Kanagawa Institute of Industrial Science and Technology.
\end{acknowledgments}

\bibliography{AFC_simulation}

\end{document}